\documentclass[runningheads]{llncs}
\usepackage{amsmath}
\usepackage{amssymb,amsfonts}
\usepackage{subcaption}
\usepackage{pdfpages}

\usepackage{tikz}
\usetikzlibrary{positioning,calc}

\usepackage{xcolor}
\definecolor{pantone7462}{HTML}{00578a}
\definecolor{pantone312}{HTML}{009DD1}
\definecolor{pantone432}{HTML}{5adbf2}
\definecolor{pantone315}{HTML}{006E89}
\definecolor{pantone3282}{HTML}{008E96}

\definecolor{change}{HTML}{0437F2}

\definecolor{pantone234}{HTML}{af0078}
\definecolor{pantone369}{HTML}{7AB51D}
\definecolor{pantone390}{HTML}{B1C800}

\usepackage{hyperref}
\usepackage{cleveref}

\usepackage{comment}

\usepackage{placeins}
\usepackage{makecell}
\usepackage{tabularx}

\usepackage{wrapfig}

\usepackage{cite}
\usepackage{graphicx}
\usepackage{textcomp}

 \usepackage{algpseudocode}
  \algrenewcomment[1]{\(\triangleright\) #1}
 \algnewcommand{\LineComment}[1]{\State \(\triangleright\) #1}

\usepackage{xspace}
\newcommand{\bs}{\;|\;}
\newcommand{\C}{\ensuremath{\mathcal{C}}\xspace}
\newcommand{\D}{\ensuremath{\mathcal{D}}\xspace}

\newcommand{\R}{\mathbb{R}}

\newcommand{\Rpz}{\mathbb{R}_{\geq 0}}

\newcommand{\N}{\mathbb{N}}

\newcommand{\Z}{\mathbb{Z}}

\usepackage{upgreek}

\usepackage{todonotes}

\newcommand{\ass}{\nu}
\newcommand{\Ass}{\mathcal{V}}

\renewcommand{\H}{\mathcal{H}\xspace}
\newcommand{\Loc}{\mathit{Loc}}
\newcommand{\Var}{\mathit{Var}} 
\newcommand{\Edge}{\mathit{Edge}}
\newcommand{\Act}{\mathit{Act}}
\newcommand{\F}{\mathcal{F}}
\newcommand{\Inv}{\mathit{Inv}}
\newcommand{\Init}{\mathit{Init}}

\newcommand{\Val}{\R^d}

\newcommand{\kernelJump}{\Psi_d}
\newcommand{\kernelDelay}{\Psi_c}

\newcommand{\resamp}[1]{\epsilon^r_{#1}}
       
\newcommand{\Edgeresampcomp}{\Edge_r^{\mathit{comp}}}
\newcommand{\Edgeresampdecomp}{\Edge_r^{\mathit{decomp}}}

\newcommand{\CS}{\C}

\newcommand{\HA}{HA\xspace}

\newcommand{\suchthat}{\,|\,}
\newcommand{\state}{\sigma}
\newcommand{\States}{\Sigma}
\newcommand{\StatesInv}{\States_{\Inv}}
\newcommand{\hpath}{\pi}
\newcommand{\Paths}{\Pi}
\newcommand{\finPaths}{\Paths_{\textit{fin}}}

\newcommand{\support}{\textit{supp}}
\newcommand{\Distr}{\textit{Dist}}
\newcommand{\DistrC}{\textit{Dist}_c}
\newcommand{\DistrD}{\textit{Dist}_d}

\newcommand{\tmax}{t_{\textit{max}}}
\newcommand{\DistJumpKernel}{\operatorname{Dist}^{\kernelJump}_{\sigma}}
\newcommand{\DistDelayKernel}{\operatorname{Dist}^{\kernelDelay}_{\sigma}}

\newcommand{\realisierungen}{\mathcal{R}}

\newcommand{\Jumps}{E_{\States}}
\newcommand{\JumpsPlus}{E_{\Loc}}

\newcommand{\dep}{\mathit{len}}

\newcommand{\dur}{\mathit{dur}}

\newcommand{\Flow}{\textit{Flow}}

\newcommand{\Times}{T}

\renewcommand{\Pr}{\textit{Pr}}
\newcommand{\proj}{\!\!\downarrow_{d}}
\newcommand{\jumpt}{t_{\textit{jump}}}

\tikzset{state/.style={ draw,text opacity=1,  minimum height=4em, minimum width=3.5em,align=center,inner sep=5pt,rectangle, rounded corners}}

\usepackage[T1]{fontenc}
\usepackage{graphicx}
 \usepackage{color}

\begin{document}

\title{Comparing Two Approaches to Include {Stochasticity} in Hybrid Automata\thanks{This work is supported by the DFG grant 471367371.}}

\author{Lisa Willemsen\inst{1}\orcidID{0000-0002-0418-9854} \and Anne Remke\inst{2\and 1}\orcidID{0000-0002-5912-4767} \and  Erika {\'A}brah{\'a}m\inst{3}\orcidID{0000-0002-5647-6134} }
\authorrunning{L. Willemsen et al.}
\titlerunning{Comparing Two Approaches to Include Stochasticity in Hybrid Automata}
\institute{University of Twente, Enschede, The Netherlands \\ 
	\email{l.c.willemsen@utwente.nl}
	\and
University of Münster, Münster, Germany \\ \email{anne.remke@uni-muenster.de}
\and
RWTH Aachen University, Aachen, Germany \\ \email{abraham@cs.rwth-aachen.de}
}
\maketitle
\begin{abstract}
Different stochastic extensions of hybrid automata have been proposed in the past, with unclear expressivity relations between them. To structure and relate these modeling languages, in this paper we formalize two alternative approaches to extend hybrid automata with sto\-chas\-tic choices of discrete events and their time points. The first approach, which we call decomposed scheduling, 
 adds stochasticity
via stochastic races, choosing random time points for the possible discrete events and executing a winner with an earliest time. In contrast, composed scheduling first samples the time point of the next event and then the event to be executed at the sampled time point. We relate the two approaches regarding their expressivity and categorize available stochastic extensions of hybrid automata from the literature.
\keywords{Formal Modelling \and Stochastic Hybrid Models  \and Classification \and
Expressivity.}
\end{abstract}

\section{Introduction}

\emph{Hybrid automata (\HA)} \cite{Henzinger1998WhatsAutomata} are well-suited to model the interplay of continu\-ous and discrete  behavior. Hybrid automata naturally exhibit non-determinism, e.g., \emph{discrete non-determinism} via  multiple simultaneously enabled discrete events (so-called jumps), or  \emph{time non-determinism} via time evolution with  non-deter\-min\-is\-tic duration.
In this paper we focus on these aspects and assume that the initial state, successor states of jumps, as well as the continuous evolution of the system state during time elapse are deterministic.

{During the execution of a hybrid automaton, every non-deterministic choice has to be resolved by a scheduler.
Hybrid automata have been extended with stochastic choices in  multiple ways, leading to \emph{stochastic hybrid models (SHM)} \cite{Lygeros2010StochasticApplications, Bertrand2014StochasticAutomata} with different features and expressivity. 
In most existing formalisms on SHM, all decisions are made randomly and  they completely replace the non-determinism present in the underlying \HA.}

For example discrete probability distributions have been added to decidable subclasses of hybrid automata~\cite{Sproston2000DecidableAutomata,Sproston2015VerificationAutomata},
which can be analyzed e.g. with the SISAT tool using abstraction \cite{Teige2009Constraint-BasedSystems}.
Another possible extension are stochastic delays via stochastic resets or random clocks \cite{Pilch2021OptimizingFlowpipe-Construction,Delicaris2023Rectangular}, which are sampled from a continuous distribution to determine how much time should pass between consecutive discrete jumps.
In the PROHVER \cite{Franzle2011MeasurabilitySystems} framework, stochastic resets are abstracted to non-deterministic probabilistic resets. Another over-approximating approach \cite{Hahn2013ASystems} discretizes
the support of the random variables and then abstracts to Markov decision processes. 
Some of these formalisms model stochastic choices by stochastic kernels \cite{Bertrand2014StochasticAutomata,Lygeros2010StochasticApplications}, each of them responsible for random decisions of a certain kind; we refer to this technique as \emph{composed scheduling}. Others use several stochastically independent random decisions that are in ``race'' \cite{Delicaris2023Rectangular,Pilch2021OptimizingFlowpipe-Construction}; we call this approach \emph{decomposed scheduling}. 
Due to these differences  and diverging mathematical notation, it is often hard to compare existing formalisms with respect to their expressivity.

\paragraph{Contribution.}
(i) In this paper, we formalize two stochastic \HA extensions, that implement composed respectively decomposed scheduling.
(ii) We define the stochastic processes induced by each of the approaches. 
(iii) We relate the expressivity of the two approaches and show that composed scheduling is more expressive than decomposed scheduling.
(iv) We discuss how existing formalisms implement such stochastic choices and relate different lines of work.

\paragraph{Outline.}  \Cref{sec:Fundamentals} introduces \HA and the necessary stochastic notation. \Cref{sec:AddingStochasticity} formalizes and relates the two \HA extensions. \Cref{sec:Relation} discusses related work and classifies existing formalisms. \Cref{sec:Conclusion} concludes the paper.

\section{Fundamentals}
\label{sec:Fundamentals}
Let $\R$ denote the set of all real numbers, $\R_{\geq 0}$ the nonnegative reals, $\N$ the natural numbers (including zero) and $\Z$ the integers. For a set $S$, $2^S$ is the set of all subsets of $S$. We start with introducing hybrid automata in \Cref{subsec:HybridAutomata} and recall some basic definitions from probability theory in \Cref{sec:prelim_prob}.

\subsection{Hybrid Automata}
\label{subsec:HybridAutomata}

\emph{Hybrid automata} extend discrete transition systems with the notion of time and continuous evolution.
In the below standard definition \cite{henzinger2000theory} we omit modeling constructs for parallel composition, as they are not central for this work.

\begin{definition}[Hybrid automata: Syntax]
  \label{def:hybrid_automaton}
  A \emph{hybrid automaton (\HA)} is a tuple $\H = (\Loc,\Var,\Flow,$ $\Inv,\Edge,\Init)$ with the following components:
  \begin{itemize}
    \item $\Loc$ is a non-empty finite set of \emph{locations} or \emph{control modes}.
    \item $\Var=\{x_1,{\ldots},x_d \}$ is a finite ordered set
      of \emph{variables}. We call $d$ the \emph{dimension}, $\nu\,{\in}\,\Val$ a \emph{valuation}, and $\state=(\ell,\nu)\in\Loc{\times}\Val = \States$ a \emph{state} of $\H$.
    \item $\Flow:\Loc\rightarrow (\Val \rightarrow \Val)$ specifies for each location its \emph{flow} or \emph{dynamics}.
    \item $\Inv:\Loc\rightarrow 2^{\Val}$ specifies an \emph{invariant} for each location. We define $\StatesInv = \{(\ell,\nu)\in\States\,|\,\nu\in\Inv(\ell)\}$.
    \item $\Edge \subseteq \Loc\times 2^{\Val} \times (\Val\rightarrow\Val)\times\Loc$ is
    a finite set of \emph{discrete transitions} or
    \emph{jumps}. For a jump
    $(\ell_1,g,r,\ell_2)\in\Edge$, $\ell_1$ and $\ell_2$ are its
    \emph{source} resp. \emph{target}
    locations, $g$ its \emph{guard}, and $r$ its \emph{reset}.
    \item $\Init:\Loc\rightarrow 2^{\Val}$ defines \emph{initial} valuations. We call a state $(\ell,\nu)\in\States$ \emph{initial} if $\nu\in\Inv(\ell)\cap\Init(\ell)$.
  \end{itemize}
\end{definition}

Executions of a hybrid automaton evolve an initial state by time steps and discrete steps. \emph{Time steps} (\emph{flows}) model continuous evolution of the variable values according to the flow condition of the current location, while satisfying the current location's invariant. When flows define constant derivatives for all variables then we talk about \emph{linear behavior}, for linear predicates (i.e., linear differential equations) about \emph{linear dynamics}, and in the case of more expressive predicates (involving e.g. polynomials or trigonometric functions) about \emph{nonlinear dynamics}.
\emph{Discrete steps} (\emph{jumps}) $(\ell,g,r,\ell')\in\Edge$ can move the control from location $\ell$ to $\ell'$ and change the valuation from $\nu\in\R^d$ to $r(\nu)\in\R^d$, assuming that the jump is \emph{enabled} in $(\ell,\nu)$, i.e. $\nu\in g$ and $r(\nu)\in\Inv(\ell')$.

\begin{definition}[Hybrid automata: Semantics]
  \noindent
  For a hybrid automaton $\H = (\Loc,\Var,\Flow,\Inv,\Edge,\Init)$ of dimension $d$, its \emph{operational semantics} is given by the following rules:
  \[
    \begin{array}[h]{r@{}l}
      \begin{array}[h]{c}
        \ell\in\Loc\quad \nu,\nu'\in\Val \quad
        t\in\R_{\geq 0} \quad
        f:[0,t]\rightarrow \Val \quad
        df/dt=\dot{f}:(0,t)\rightarrow\Val\\
        f(0)=\nu\quad f(t)=\nu'\qquad                                                
        \forall t'\in(0,t).\ \dot{f}(t')=\Flow(\ell)(f(t'))\\
        \forall t'\in[0,t].\ f(t')\in\Inv(\ell)
        \\\hline
        (\ell,\nu)\stackrel{t}{\rightarrow} (\ell,\nu')
      \end{array}
       & \begin{array}{c} \\ \\ 
        \ \texttt{Flow}
      \end{array}
    \end{array}
  \]
  \[
    \begin{array}[h]{r@{}l}
      \begin{array}[h]{c}
        e=(\ell,g,r,\ell')\in\Edge\quad
        \nu,\nu'\in\Val \quad
        \nu\in g\quad
        \nu'=r(\nu)\quad
        \nu'\in\Inv(\ell') \\\hline
        (\ell,\nu)\stackrel{e}{\rightarrow} (\ell',\nu')
      \end{array}
       & \ \texttt{Jump}
    \end{array}
  \]
\end{definition}

\noindent Let $\H = (\Loc,\Var,\Flow,\Inv,\Edge,\Init)$ be a \HA of dimension $d$. 
A \emph{path} of $\H$ is a finite or infinite sequence  $\hpath=\state_0\xrightarrow{t_0} \state_0'\xrightarrow{e_0}\state_1\xrightarrow{t_1}\ldots$ of alternating time steps and jumps with $\state_0\in\StatesInv$;  $\hpath$ is said to be \emph{initial} if $\state_0$ is initial,
we define its \emph{length} $\dep(\hpath)$ as the number of jumps in it, and its \emph{duration} $\dur(\hpath)$ as the sum of the durations of all of its time steps.
Let $\Paths(\state)$  and $\finPaths(\state)$ be the set of all infinite resp. finite paths starting in $\state\in\StatesInv$. 
A state $\state\in\States$ of $\H$ is \emph{reachable} iff there is an initial path leading to it. 

An infinite path is \emph{time-convergent} if its duration is finite, and \emph{time-divergent} otherwise. A state of $\H$ has a \emph{timelock} iff all infinite paths starting in it are time-convergent; $\H$ has a \emph{timelock} if any of its reachable states has a timelock, and is \emph{timelock-free} otherwise. An infinite path is \emph{Zeno} iff it is time-convergent and contains infinitely many jumps.  $\H$ is \emph{Zeno-free} if it has no initial Zeno path.

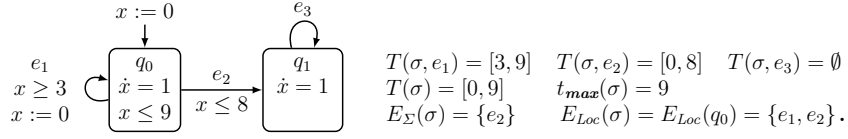
\begin{figure}[t]
\centering
\scalebox{0.7}{
  \begin{tikzpicture}[
	node distance=6cm,
	on grid,
	thick,
	font=\bfseries\large]

\node
[state] (q0) at (0,0)
{
	$ q_0$   \\
	$	\dot{x}=1$ \\
		$x\leq 9 $
};

\node[state] (q1) at (3,0)
{
	$ q_1$   \\
	$\dot{x}=1 $ \\
	\phantom{$x\leq 9$ }
};

\node[draw=none, align=center,  above= 1.5cm of q0] (init) {$x:=0$};

\path [-latex] (init.south) edge [] ($(q0.north)+(0,0)$)
(q0.east) edge []  node[above] {$e_2$} node[below] {$x\leq 8$}  ($(q1.west)+(0,0)$)
(q1) edge[loop above,looseness=4] node[above] {$e_3$} (q1)
(q0) edge[loop left,looseness=4] node[align=center, left=0.2]{$e_{1}$\\$x\geq3$\\$x:=0$}(q0)
;

\node at (9,0) {
  $\begin{array}{l@{\quad}l@{\quad}l}
    \Times(\state, e_1)=[3,9] &
    \Times(\state, e_2)=[0,8]&
    \Times(\state, e_3)=\emptyset \\
    \Times(\state)=
             [0,9]&
    \tmax(\state)=9  \\
    \Jumps(\state)=\{e_2\}&
    \multicolumn{2}{l}{\JumpsPlus(\state)=\JumpsPlus(q_0)=\{e_1,e_2\}}.
\end{array}$};

\end{tikzpicture}
}
     \caption{A hybrid automaton and characteristics of its state $\sigma=(q_0,\ass)$, $\ass(x)=0$.}
     \label{fig:exampleSets}
\end{figure}

We discuss how choices over jumps and the length of time steps are made stochastically in \Cref{sec:AddingStochasticity},
where we use the following notions, partly generalized to hybrid automata from   \cite{Bertrand2014StochasticAutomata}  and illustrated in \Cref{fig:exampleSets}. For $ e\in \Edge $ and  $ \state\in\StatesInv$, we define
\[
\begin{array}{l@{\,}cl@{\ \ }l}
	\Times(\state ,e) & =& \{t\in\Rpz \suchthat \exists \state',\state''\in\StatesInv.\ \state {\xrightarrow{t}}\state'\wedge \state'{\xrightarrow{e}}\state''\},&
	\Times(\state )=\bigcup_{e\in \Edge} \Times(\state ,e),\\
	\tmax(\state)&=&\operatorname{sup}\{t\in\R\suchthat \exists \state'\in\StatesInv.\ \state \xrightarrow{t}\state'\},\\

	\Jumps(\state)&=&\{e\in \Edge \suchthat \exists \state'\in\StatesInv.\ \state\xrightarrow{e}\state'\},\\
	\JumpsPlus(\ell)&=&\cup_{\ass\in\Ass}\, \Jumps((\ell,\ass)),&
        \JumpsPlus((\ell,\ass))=\JumpsPlus(\ell).
\end{array}
\]
\noindent We call $\state$ \emph{jump-enabled}\footnote{In the original definition, this is called \emph{non-blocking}.} iff $ \Times(\state){\not=}\emptyset $, and \emph{immediate-jump-enabled} iff $\Jumps(\state)\not=\emptyset$. 

\subsection{Basic Stochastic Notions}
\label{sec:prelim_prob}

A random \emph{experiment} has an uncertain outcome from a \emph{sample space} $\Omega$, whose subsets are called \emph{events}.
A \emph{$\sigma$-algebra} $\F$ is a set of events containing the maximal event $\Omega$ and being closed under complement and countable union. The standard Borel $\sigma$-algebra  $\mathcal{B}(\Omega)$  is the smallest $\sigma$-algebra containing all open events.
An event is \emph{$\F$-measurable} if it is in $\F$.
The pair $(\Omega, \F)$ is called a \emph{measurable space}.

Given $(\Omega, \F)$, a \emph{probability measure} is a function $\Pr : \F \rightarrow [0, 1] \subseteq \R$ with
(i) $\Pr (\Omega) = 1$,
(ii) $\Pr(E)=1-\Pr(\bar{E})$ for all $E \in \F$ and 
(iii) $\Pr (\bigcup_{i=0}^{\infty} E_i ) = \Sigma_{i=0}^{\infty} \Pr (E_i )$ for any $E_i \in  \F$ with $E_i \cap E_j = \emptyset$ for all $i, j \in \N$, $i \not= j$.

A \emph{probability space} is a triple $(\Omega, \F, \Pr )$ with
$(\Omega, \F)$ a measurable space and
$\Pr$ a probability measure for $(\Omega, \F)$.

Let $(\Omega, \F)$ and $(S, \Sigma)$ be measurable spaces,
$X : \Omega {\rightarrow} S$, $s {\in} S$ and $\sigma{\in}\Sigma$. We define
$X \sim s$ to be $\{\omega \in\Omega\suchthat X (\omega) \sim s\}$ and
    $X^{-1}(\sigma) = \bigcup_{s\in\sigma} (X \sim s)$ with $\sim\in\{\leq,<,=,>,\geq\}$.
$X$ is \emph{measurable} (wrt. $(\Omega, \F)$ and $(S, \Sigma)$) if $X^{-1}(\sigma) \in \F$ for all $\sigma \in\Sigma$.
A \emph{random variable} is a measurable function $X : \Omega \rightarrow S$;
we call $X(\omega)$ the \emph{realization} of $X$ for $\omega\in\Omega$.

For the following we instantiate $\Omega=S=\Rpz$, $\F=2^{\Rpz}$ and $X$ the identity.  For $f : \Rpz \to \Rpz$ we define its \emph{support} as $\support(f) = \{\omega \in \R \suchthat f (\omega) > 0\}$, with: 
\begin{itemize}
\item If $\support(f)$ is countable and $\sum_{\omega\in \support(f)} f (\omega) = 1$,  $f$ is called a
  \emph{discrete probability distribution}, which induces the unique probability measure $\Pr:2^{\Rpz}\rightarrow[0,1]$ with $\Pr(E)=\sum_{\omega\in E\cap \support(f)} f(\omega)$ for all $E\subseteq \Rpz$. 
\item If $f$ is absolute continuous with $\int_{0}^{\infty}  f(\omega) \textit{d}\omega = 1$ then $f$ is called a
  \emph{continuous probability distribution} or a \emph{probability density function (PDF)}, which induces for all $a\in \Rpz$ the unique probability measure $\Pr:2^{\Rpz}\rightarrow [0,1]$ with $\Pr(X\leq a)=\int_{0}^{a} f(\omega) \textit{d}\omega$, and the \emph{cumulative distribution function} (CDF) $F:\Rpz\to[0,1]$ with $F(a)=\Pr(X\leq a)$.
\end{itemize}

We denote the set of all discrete resp. continuous probability distributions by $\DistrD$ resp. $\DistrC$ and call elements from $\Distr=\DistrD\cup\DistrC$ \emph{probability distributions}. A random variable is \emph{discrete} if its underlying probability measure $\Pr$ is induced by a discrete probability distribution, and \emph{continuous} otherwise. 
By $\mathbb{X}_d$ and $\mathbb{X}_c$ we denote the set of all discrete resp. continuous random variables. 

\label{def:StochasticKernel}
Given two measurable spaces $ (\Omega_1,\F_1) $ and  $ (\Omega_2,\F_2) $, a \emph{stochastic kernel from $ (\Omega_1,\F_1) $ to $ (\Omega_2,\F_2) $} \cite{Klenke2014ProbabilityCourse} is a function $\kappa:\F_2 \times \Omega_1\to[0,1]$ with:
	\begin{itemize}
	\item For each $ E_2\in\F_2 $, the function $f_{E_2}^{\kappa}: \Omega_1\rightarrow [0,1]$ with $f_{E_2}^{\kappa}(\omega_1)=\kappa(E_2,\omega_1) $ is measurable w.r.t. $(\Omega_1, \F_1)$ and $([0,1],\mathcal{B}([0,1])$.
	\item For  each $ \omega_1\in\Omega_1 $, the function $ \Pr^{\kappa}_{\omega_1}: \F_2 \rightarrow [0,1]$ with $\Pr^{\kappa}_{\omega_1}(E_2)=\kappa(E_2,\omega_1) $ is a probability measure on $(\Omega_2,\F_2)$.
	\end{itemize}
        Stochastic kernels are used to express the state-dependent probability $\kappa(E_2,\omega_1)$ of event $E_2\in\F_2$ in system state $\omega_1\in\Omega_1$.  $\kappa$ is  \emph{discrete} if each $\Pr^{\kappa}_{\omega_1}$ can be induced by a discrete probability distribution, and \emph{continuous} otherwise.

        A \emph{stochastic process} over an index set $T$ is a family of random variables $\{X(t) \bs t\in T\}$, defined over a common probability space and taking values in the same measurable space.
In this work we use continuous-time stochastic processes with $T\subseteq\Rpz$. A stochastic process has the \emph{Markov property} if its future is independent of its past evolution \cite{Klenke2014ProbabilityCourse}.

\section{Extending Hybrid Automata with Stochasticity}
\label{sec:AddingStochasticity}
\label{sec:probextensions}

In this section we formalize two stochastic \HA extensions: \emph{Composed scheduling}, introduced in \Cref{subsec:locationBasedScheduling}, randomly chooses first a delay and then a jump to be taken after the delay.  Conversely, \emph{decomposed scheduling}, introduced in \Cref{subsec:decomposed}, chooses a delay for each jump separately, where the jump with the minimal delay is taken.
Our approach chooses delays from $\Rpz$, which might be unrealizable due to invariants, or after which no jump might be enabled, so we need to introduce mechanisms to manage these cases. We mention that other formalisms like \cite{Bertrand2014StochasticAutomata} assume an ``oracle'' and can therefore sample only over realizable time delays after which there is an enabled jump.

To put a clear focus on the differences between composed and decomposed scheduling, our \HA definition assumes deterministic flows and jump resets. In addition, for simplicity we assume in the following a unique initial state. However, our languages and results can easily be extended to relax these restrictions.

In \Cref{subsec:decomposed,subsec:locationBasedScheduling} we assume that all invariants evaluate to true; we use $\top$ to denote the trivial invariant, i.e. $\top(\ell)=\R^d$ for all $\ell\in\Loc$.  In \Cref{subsec:invariants} we discuss which adaptions in the definitions are required to apply (de-)composed scheduling to \HA with invariants.

\subsection{Composed Scheduling}
\label{subsec:locationBasedScheduling}
In \emph{composed scheduling}, the durations of time steps and the jumps to be taken are sampled according to two stochastic kernels $\kernelDelay$ resp. $\kernelJump$.
For a state $\sigma$, the probability distributions induced by $\kernelDelay$ and $\kernelJump$ are denoted $\DistDelayKernel$ resp. $\DistJumpKernel$.

To schedule time delays, approaches like e.g. \cite{Bertrand2014StochasticAutomata} sample only durations after which there exists an enabled jump. This alternative is meaningful for decidable subclasses only (as one needs to determine valid time durations) and for them it would result in the same expressivity as our approach, which is as follows.

Assume a $d$-dimensional hybrid automaton $\mathcal{H}_{\textit{in}}=(\Loc, \Var, \Edge_{\textit{in}}, \Act, \allowbreak\top, \Init)$.
Without an oracle, it might happen that $\kernelDelay$ samples a time duration after which there are no enabled jumps. To handle such cases, we introduce for each location $\ell\in\Loc$ a unique \emph{(composed) resampling jump} $\resamp{\ell}=(\ell,g,r,\ell)$ with guard $g=\R\backslash (\cup_{(\ell,g',r',\ell')\in \Edge_{\textit{in}}}g')$ and reset $r(\nu)=\nu$ for all $\nu\in\R^d$. Let $\Edgeresampcomp=\{\resamp{\ell}\,|\,\ell\in\Loc\}$. We call $\mathcal{H}=(\Loc, \Var, \Edge_{\textit{in}}\cup\Edgeresampcomp, \Act, \allowbreak\top, \Init)$ the \emph{composed resampling extension} of $\mathcal{H}_{\textit{in}}$.

\begin{definition}[Composed Syntax]
\label{def:TALocationBased}
A \emph{hybrid automaton with composed scheduling} is a tuple $\mathcal{C}=(\mathcal{H}, \kernelDelay,\kernelJump)$, where: 
\begin{itemize}
\item   $\mathcal{H}=(\Loc, \Var, \Edge, \Act, \allowbreak\top, \Init)$ with states $\States$ is the composed resampling extension of a \HA $\mathcal{H}_{\textit{in}}$ with trivial invariants and deterministic initial state.
\item $\kernelDelay:\mathcal{B}(\Rpz)\times\States\to [0,1]$ is a continuous  stochastic kernel from $(\States,\allowbreak \mathcal{B}(\States))$ to $(\Rpz,\mathcal{B}(\Rpz))$.
\item $\kernelJump:\mathcal{B}(\Edge)\times\States \to [0,1]$ is a discrete stochastic kernel from $(\States, \allowbreak \mathcal{B}(\States))$ to $(\Edge,\mathcal{B}(\Edge))$ such that $\support(\DistJumpKernel)\subseteq \Jumps(\sigma)$ for all $\sigma\in\States$. 
\end{itemize}
\end{definition}
After each time step, if any non-resampling jump of $\mathcal{H}_{\textit{in}}$ is enabled then resampling is scheduled with probability $0$, and otherwise with probability $1$. In each jump successor state $\sigma$, a fresh delay is sampled according to $\operatorname{Dist}^{\kernelDelay}_{\sigma}$.

\begin{definition}[Composed Semantics]
\label{def:SemanticComposed}
Assume a HA with composed  scheduling $\CS=(\H, \kernelDelay,\kernelJump)$. A \emph{path of $\CS$} is a path $\pi = \sigma_0 \xrightarrow{t_0} \sigma_0' \xrightarrow{e_0} 
\sigma_1 \xrightarrow{t_1} \ldots $ of $\H$ 
with $t_j\in\support(\operatorname{Dist}^{\kernelDelay}_{\sigma_{j}})$ and  $e_j\in\support(\operatorname{Dist}^{\kernelJump}_{\sigma_{j}'})$ for all $j\geq 0$.
\end{definition}
\subsection{Decomposed Scheduling}
\label{subsec:decomposed}

Let $\mathcal{H}_{\textit{in}}=(\Loc, \Var, \Edge_{\textit{in}}, \Act, \allowbreak\top, \allowbreak \Init)$ be a $d$-dimensional \HA. Instead of centralized decisions via stochastic kernels,  \emph{decomposed scheduling}
 chooses jumps and the length of time steps
by associating each jump with a continuous random variable from a non-empty set $\mathbb{X}=\{X_1,\ldots,X_k\}$ via a function $\operatorname{a}_{\textit{in}}:\Edge_{\textit{in}} \to\{1,\ldots,k\}$, such that two jumps with the same random variable are never enabled simultaneously. We call such a pair $(\H_{\textit{in}},\operatorname{a}_{\textit{in}})$ an \emph{$\mathbb{X}$-labeled \HA}.

The realisations of the random variables indicate the delay after which a jump with the given random variable should be taken; if no such jump is enabled then we again need a mechanism for resampling.

For each location $\ell\in\Loc$ and  random variable $X_i\in \mathbb{X}$ we introduce a  \emph{(decomposed) resampling jump} $\resamp{\ell,i}=(\ell,g_{\ell,i},r,\ell)$ with reset $r(\nu)=\nu$ for all $\nu\in\R^d$ and guard $g_{\ell,i}=\R^d\setminus(\cup_{e\in\{e'=(\ell,g,r,\ell')\in\Edge_{\textit{in}}\,|\, \operatorname{a}_{\textit{in}}(e')=i\}} \ g)$.
We extend the edge set to $\Edge=\Edge_{\textit{in}}\cup\Edgeresampdecomp$ with the resampling edges $\Edgeresampdecomp=\{\resamp{\ell,i}\,|\,\ell\in\Loc\wedge 1\leq i\leq k\}$. Let  $\mathcal{H}=(\Loc, \Var, \Edge, \Act, \allowbreak\top, \Init)$.

We extend also $\operatorname{a}_{\textit{in}}$ to cover the resampling edges, defining $\operatorname{a}:\Edge \to\{1,\ldots,k\}$ with $\operatorname{a}(e)=\operatorname{a}_{\textit{in}}(e)$ for $e\in\Edge_{\textit{in}}$ and $\operatorname{a}(\resamp{\ell,i})=i$ for $\resamp{\ell,i}\in \Edgeresampdecomp$. Let
 $\operatorname{a}^{-1}:\{1,\ldots,k\}\to 2^{\Edge}$ with  $\operatorname{a}^{-1}(i)=\{e\in\Edge\,|\,\operatorname{a}(e)=i\}$ for $i=1,\ldots,k$.

We call $(\H,\operatorname{a})$ the \emph{decomposed resampling extension} of the $\mathbb{X}$-labeled \HA $(\mathcal{H}_{\textit{in}},\operatorname{a}_{\textit{in}})$. Note that $(\H,\operatorname{a})$ itself is an $\mathbb{X}$-labeled \HA.

\begin{definition}[Decomposed Syntax]\label{def:stoch-race-sched}
  A \emph{hybrid automaton with  decomposed scheduling} is a tuple $\mathcal{D}=(\mathcal{H},\mathbb{X},\operatorname{a})$ where:
  \begin{itemize}
  \item $\mathbb{X}=\{X_1,\ldots,X_k\}$ is a finite non-empty ordered set of continuous random variables.
  \item $(\H,\operatorname{a})$ with  $\mathcal{H}=(\Loc, \Var, \Edge, \Act, \allowbreak\top, \Init)$ is the decomposed resampling extension of some $\mathbb{X}$-labeled \HA $(\mathcal{H}_{\textit{in}},\operatorname{a}_{\textit{in}})$, where  $\mathcal{H}_{\textit{in}}$ has trivial invariants and a deterministic initial state.
      \end{itemize}
\end{definition}

We store the current realizations of the random variables in a sequence $\realisierungen=(x_1,\ldots,x_k)\in\Rpz^{k}$, and use $\realisierungen[j]$ to refer to $x_j$. The stochastic race between the random variables  is ``won'' by the random variable which \emph{expires} first as it has a smallest current realisation.
The presence of resampling jumps ensures, that a jump can be scheduled, even if there is no enabled edge associated with the winning random variable.
Note that, since all random variables follow a continuous probability distribution, the probability that two random variables expire at the same time is $0$ and in this case it is irrelevant which jump is taken.

\begin{definition}[Decomposed Semantics]
  \label{def:SemanticDeomposed}
        Assume a \HA with decomposed scheduling $\mathcal{D}=(\H,\mathbb{X},\operatorname{a})$ with $\mathbb{X}=\{X_1,\ldots,X_k\}$. A \emph{path of $\mathcal{D}$} has the form $\pi= (\sigma_0,\realisierungen_0) \xrightarrow{t_0} (\sigma_0',\realisierungen'_0)\xrightarrow{e_0} (\sigma_1,\realisierungen_1) \xrightarrow{t_1}\ldots$ with $\sigma_i=(\ell_i,\nu_i)$, $\sigma_i'=(\ell_i,\nu_i')$ such that $\sigma_0 \xrightarrow{t_0} \sigma_0'\xrightarrow{e_0} \sigma_1 \xrightarrow{t_1}\ldots$        
        is a path of $\H$ and such that for all $i\in\N$: 
\begin{itemize}
  \item $\realisierungen_i,\realisierungen_i'\in\Rpz^k$ and for all $j\in\{1,\ldots,k\}$, $\realisierungen_0[j]$ is sampled according to $X_j$'s probability distribution in $\sigma_0$.
  \item $t_i=\min_{j\in\{1,\ldots,k\}}\realisierungen_i[j]$ and $\realisierungen_i'[j]=\realisierungen_i[j]-t_i$ for all $j\in\{1,\ldots,k\}$.
  \item $\realisierungen_i'[m_i]=0$ for $m_i=\operatorname{a}(e_i)$.
  \item The value $\realisierungen_{i+1}[m_i]$ is sampled according to $X_{m_i}$'s probability distribution in $\sigma_{i+1}$, and $\realisierungen_{i+1}[j]=\realisierungen_i'[j]$ for all $j\in\{1,\ldots,k\}\setminus\{m_i\}$.
\end{itemize}
\end{definition}

\begin{figure}[tb]
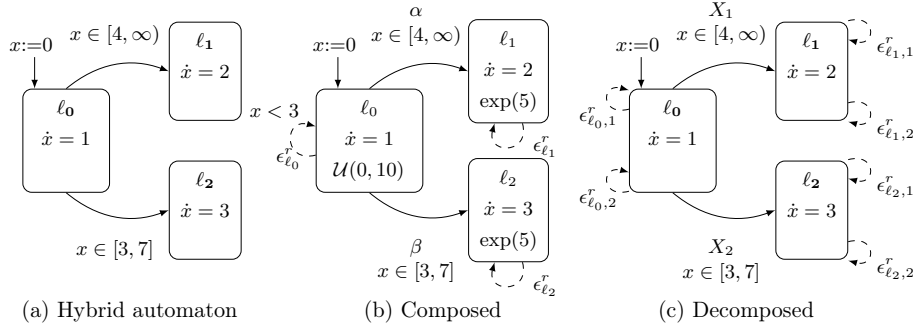
	
	\scalebox{0.84}{
		\begin{tikzpicture}
			\begin{scope}
				\input{img/NonDet.tex}
			\end{scope}
   
			\begin{scope}[xshift=4.8cm]
				\input{img/Composed.tex}
			\end{scope}
			
			\begin{scope}[xshift=9.6cm]
				\input{img/Decomposed.tex}
			\end{scope}
	\end{tikzpicture}}
	\caption{Illustration to  \Cref{example:CompareSemantics}.
 Hybrid automaton shown in (a) is extended with either composed scheduling in (b) or decomposed scheduling in (c).}
	\label{fig:ExampleCompDecomp}
\end{figure}

\begin{example}[(De-)composed scheduling]
    \label{example:CompareSemantics}
  We illustrate the application of composed and decomposed scheduling on the example \HA depicted in  \Cref{fig:ExampleCompDecomp}(a). Note that the resulting \HA with (de-)composed scheduling in this example have different underlying stochastic processes (c.f. \Cref{sec:induced}).
  
In the composed scheduling in \Cref{fig:ExampleCompDecomp}(b),
choices over the time steps' durations are governed by a kernel $\kernelDelay$,
which characterises for each state with location $\ell_0$ a uniform distribution $\operatorname{Dist}^{\kernelDelay}_{(\ell_0,\ass)} =\mathcal{U}(0,10)$, and an exponential distribution $\operatorname{Dist}^{\kernelDelay}_{(\ell_i,\ass)} =\exp(5)$ otherwise. 
The kernel $\kernelJump$  characterises a discrete probability distribution over enabled jumps. Hence, $\kernelJump(e_1,(\ell_0,\ass))=\alpha$, $\kernelJump(e_2,(\ell_0,\ass))=\beta $ and $\kernelJump(\resamp{\ell_0},(\ell_0,\ass))=1-\alpha-\beta$, where $e_1$ denotes the jump to location $\ell_1$ and $e_2$  the jump to location $\ell_2$. The values of $\alpha$ and $\beta$ depend on the enabling status of $e_1$ and $e_2$. Hence, $\alpha=1 $ and $ \beta=0$, if $e_1$ is the only jump enabled. Similarly, $\alpha=0$, $\beta=1$ in case $e_2$ is enabled but $e_1$ is not. If both jumps are enabled then we set $\alpha=0.7$ and $\beta=0.3$. If neither $e_1$ nor $e_2$ is enabled, then $\alpha=\beta=0$ and the resampling edge is scheduled with probability $1$.

In the decomposed scheduling in \Cref{fig:ExampleCompDecomp}(c),
there is a race between two competing random variables $X_1$ and $X_2$, and the "winner" decides on the delay as well as on the scheduled jump. For our example, in the initial state $\sigma_0=(\ell_0, \ass)$ with $\ass(x)=0$ we sample the values $x_1$ and $x_2$ for $X_1$ resp. $X_2$ from the exponential distribution with parameter $0.2$. 
After a delay of $t=\operatorname{min}\{x_1,x_2\}$ a jump takes place. If $t=x_1$ then it is the unique enabled jump associated with $X_1$ (i.e. either the jump to $\ell_1$ or the resampling jump $\epsilon_{\ell_0,1}$), otherwise if  $t=x_2$ then it is the unique enabled jump associated with $X_2$ (i.e. either the jump to $\ell_2$ or the resampling jump $\epsilon_{\ell_0,2}$).
\end{example}

\subsection{Induced Stochastic Process} 
\label{sec:induced}
The semantics of a hybrid automaton with unique initial state and composed or decomposed scheduling is fully stochastic. The execution  corresponds to a continuous-time stochastic
process $\{X(t) \bs t \in \Rpz\}$, where the random variable $X(t)$ takes values from the measurable space
$(\States, \mathcal{B}(\States))$.

The corresponding probability space for finite paths of length $\dep(\hpath)=n$ is given by $(\Omega, \mathcal{F}, \Pr)$, where $\Omega =
(\States)^n$ and $\mathcal{F} = \mathcal{B}(\Omega)$. 
The probability measure $\Pr$ also  depends on the chosen scheduling method. Since the probability of a single path is in most cases zero, we define measurable probabilities for \emph{traces}.

\begin{definition}[Trace]
  A \emph{trace}  of a \HA $\H$ is a finite sequence $\tau = (\sigma_0, e_0, e_1,\allowbreak \ldots, e_{n})$ composed of a state $\sigma_0\in\StatesInv$ and a (possibly empty) sequence of jumps $e_0,\dots,e_{n}$ of $\H$. The trace $\tau$ is \emph{initial} if $\sigma_0$ is an initial state of $\H$.
  A \emph{sub-trace}  of  $\tau $ is a trace $\tau'=(\sigma_i,e_i,\ldots,e_n)$ for some $1\leq i\leq n$,
 where $\sigma_i$ is reachable in $\H$ through a path $\state_0\xrightarrow{t_0} \state_0'\xrightarrow{e_0}\state_1\xrightarrow{t_1}\ldots\xrightarrow{e_{i-1}}\state_{i}$.

\noindent A \emph{trace} of a \HA with (de-)composed scheduling is a trace of its underlying \HA.
 \end{definition}
The probability measure of a trace is obtained recursively by  integrating over the enabling time of the first jump and taking into account the corresponding jump probabilities.
Note that traces of $\mathcal{C}$ might include resampling jumps.

\begin{definition}[Composed Probability Measure]
  For a HA with composed scheduling $\mathcal{C}=(\H,\kernelDelay,\kernelJump)$ and a trace  $\tau = (\sigma_0,\allowbreak  e_0,e_1 \ldots, e_{n})$ of $\mathcal{C}$
  we define $\Pr_\mathcal{C}(\tau)$ to be $1$ if $\tau = (\sigma_0)$, and otherwise 
\begin{align}
\label{eq:problazy}
&\Pr_\mathcal{C}(\tau) =
\int_{t\in  \Times(\sigma_0,e_0)}\kernelDelay(\sigma_0)(t)\cdot \kernelJump(\sigma_0')(e_0)\cdot \Pr_\mathcal{C}(\sigma_1, e_1, \ldots, e_n) dt
\end{align}
where $\sigma_0'$ and $\sigma_1$ are the unique states of $\H$ with $\sigma_0\xrightarrow{t}\sigma_0'$ and $\sigma_0'\xrightarrow{e_0}\sigma_1$.
\end{definition}
\begin{definition}[Decomposed Probability Measure]
 Assume a \HA with decomposed scheduling $\mathcal{D}=(\mathcal{H},\mathbb{X},a)$, $\mathbb{X}=\{X_1, X_2, \dots , X_k\}$. For any state $\sigma$ of $\H$ and any $t\in\Rpz$, let $\sigma^t$ and be the unique state of $\H$ with $\sigma\xrightarrow{t}\sigma^t$, and for any  $e\in\Jumps(\sigma)$ let $\sigma^e$ and be the unique state of $\H$ with $\sigma\xrightarrow{e}\sigma^{e}$.

  \noindent For  a trace $\tau = (\sigma_0, e_0, e_1, \allowbreak \ldots, \allowbreak e_{n})$ of $\mathcal{D}$ we define 
\begin{eqnarray}
        \label{eq:probstochrace}
        &&\Pr_\mathcal{D}(\tau)= \\
        &&\int_{0}^{\infty} \textit{Dist}_{X_1}(\sigma_0,t_1) \ldots \int_{0}^{\infty}  \textit{Dist}_{X_k}(\sigma_0,t_k)  \
         P(\sigma_0^{\delta_0},\realisierungen,e_0, e_1 \ldots, e_n)\ dt_k \ldots dt_1 \nonumber
\end{eqnarray}
    where $\delta_0=\operatorname{min}\{t_m\bs 1 \leq m \leq k \}$, $\realisierungen\in\Rpz^k$ with $\realisierungen[m]= t_m-\delta_0$ for all $1\leq m\leq k$, and where $P$ is defined as follows.

    \noindent  For any a trace $\tau = (\sigma_0, e_0, e_1, \allowbreak \ldots, \allowbreak e_{n})$ of $\mathcal{D}$ and any $\realisierungen\in\Rpz^k$ with $\sigma_0=(\ell_0,\nu_0)$ and $m_0=\operatorname{a}(e_0)$ we set $P(\sigma_0,\realisierungen)=1$ and for a non-empty sequence $e_0,\ldots,e_n$:
\begin{eqnarray}
\label{eq:probP}
&&P(\sigma_0,\realisierungen, e_{0},\ldots,e_n) =\\ \nonumber
&&\left\{\begin{array}{l}
0
\hfill
\textit{if } e_{0}\notin\Jumps(\sigma_0) \textit{ or } \realisierungen[m_0]\not=0\\
\int_{0}^{\infty} \textit{Dist}_{X_{m_0}}(\sigma_0^{e_0}, t_{m_0})  \cdot   P((\sigma_0^{e_0})^{\delta_1}, \realisierungen', e_{1}, \ldots, e_n )  d t_{m_0}
\hspace*{2cm}
\textit{otherwise},
\end{array}\right.
\end{eqnarray}
where $\delta_1=\textit{min}(\{t_{m_0}\}\cup\{\realisierungen[m]\,|\,1\leq m\leq k\wedge m\not=m_0\})$,
$\realisierungen'[m_0]=t_{m_0}-\delta_1$ and $\realisierungen'[m]=\realisierungen[m]-\delta_1$ for all $m\in\{1,\ldots,k\}\setminus\{m_0\}$.
\end{definition}

When decomposed scheduling is applied, the above-defined probability measure over traces (which might also include resampling jumps) is obtained by sampling all random variables from their corresponding probability distributions. Afterwards, it recursively computes the probability measure $P$ of the trace for the sampled durations, after letting time elapse by the minimum realisation $\delta_0$ under all random variables.
In the definition of $P$ in Equation~\ref{eq:probP}, the first case applies if the jump $e_0$ is either disabled in the trace's starting state or the realisation of its random variable is not yet expired (i.e. not 0); in this case, the probability of the trace is $0$. Otherwise, we take the jump $e_0$ with successor state $\sigma_0^{e_0}$, resample the random variable $X_{m_0}$ of $e_0$, let again time elapse with the minimum realisation $\delta_1$ over all random variables, and apply the definition recursively. Note that $e_0$ might be a resampling jump and that though several realisations can expire simultaneouly, it might happen only with probability $0$.

\begin{remark}
    \label{req:del}
   Hybrid automata with composed scheduling $\C$ and HA with de\-composed scheduling $\D$ both extend the jump set of their underlying hybrid automaton $\H$. Therefore, $\C$ and $\D$ have more paths than $\H$. This means that Zeno paths are inherited from $\H$ to $\C$ and $\D$, however, Zeno-freedom of $\H$ does not imply Zeno-freedom of $\D$ and $\C$. 

Consider for example the hybrid automaton $\H$ from \Cref{fig:ExampleCompDecomp}(a) and its composed scheduling extension $\C$ from \Cref{fig:ExampleCompDecomp}(b). The automaton $\H$ is Zeno-free but $\C$ does have Zeno paths, e.g. all paths that take the resampling jump $\epsilon_{\ell_0}$ of $\ell_0$ infinitely often. Even though the stochastic kernel of $\C$ almost surely excludes such paths (i.e. if $\tau_k$ is the trace with $k$ repeated $\epsilon_{\ell_0}$-jumps from the initial state $\sigma_0$ then $\lim_{k\rightarrow\inf}P(\tau_k)=0$), changing the distribution in $\ell_0$ from $\mathcal{U}(0,10)$ to $\mathcal{U}(0,\frac{1-x}{2})$ would increase their probability to $1$. Hence, modelers should carefully choose stochastic distributions in order to ensure that additional Zeno behavior is almost surely excluded.
\end{remark}

\subsection{Relation of Composed and Decomposed Scheduling}
\label{subsec:relationComposed}
Previously, we discussed two different approaches on how hybrid automata can be extended with 
stochasticity. 
In this section we show that composed scheduling is \emph{more expressive} than decomposed scheduling w.r.t. \emph{Pr-equivalence}. 

\begin{definition}[Trace Probability Equivalence]
\label{def:PrEquiv}
Let $\mathcal{D}$ be a \HA with decomposed scheduling, $\mathcal{C}$ a \HA with composed scheduling, and
  $\tau$ a common trace of $\mathcal{D}$ and $\mathcal{C}$.  The trace $\tau$ is \emph{Pr-equivalent} in $\mathcal{D}$ and $\mathcal{C}$ iff $\Pr_{\mathcal{D}}(\tau) = \Pr_{\mathcal{C}}(\tau)$ and for each $\sigma$ being either the first state of $\tau$ or the first state of a sub-trace of $\tau$ if holds for all $t\in \Rpz$ that $\Pr^\sigma_{\mathcal{D}}(X\leq t)=\Pr^{\sigma}_{\mathcal{C}}(X\leq t)$, where $X$ models the duration of a time step starting in $\sigma$.

$\mathcal{D}$ and $\mathcal{C}$ are \emph{Pr-equivalent} if the sets of their initial traces are equal and each of their initial traces is Pr-equivalent.
  
\end{definition}
\begin{theorem}[Expressivity Composed vs Decomposed Scheduling]
	\label{theo:Expressivity}
\begin{enumerate}
    \item Let $\mathcal{D}$ be a \HA with decomposed scheduling. Then there is a  \HA  with composed scheduling $\mathcal{C}$ such that $\mathcal{D}$ and $\mathcal{C}$ are Pr-equivalent.
    \item  There exists a  \HA with composed scheduling $\mathcal{C}'$   such that there is no \HA with decomposed scheduling $\mathcal{D}'$  such that $\mathcal{D}'$ and $\mathcal{C}'$ are Pr-equivalent.
\end{enumerate}
\end{theorem}
\paragraph{Proof (\Cref{theo:Expressivity})}
\emph{Statement 1.}
Assume a \HA with decomposed scheduling $\mathcal{D}{=}(\mathcal{H},\mathbb{X},\operatorname{a})$ with $\mathcal{H}=(\Loc,\Var,\Flow,\allowbreak\top,\Edge,\Init)$,
$\Loc=\{\ell_1,\ldots,\ell_n\}$,
$\Var=\{v_1,\ldots,v_d\}$ and $\mathbb{X}=\{X_1,\ldots,X_k\}$. We construct a \HA  with composed scheduling $\mathcal{C}=(\mathcal{H}', \kernelDelay, \kernelJump )$ with $\mathcal{H}'=(\Loc,\Var',\Flow',\allowbreak \top,\Edge',\Init')$ as follows.

We encode into the state of $\mathcal{H}'$ for each random variable (i) the state in which it was sampled the last time and (ii) the time which has evolved since then. We also encode (iii) the time spent in the current location since last entering.

To account for (i), we introduce fresh variables $D=\{d_i\bs 1\leq i\leq k \}$ to store the location components and $ C=\{c_{i,j}\bs 1\leq i\leq k, 1\leq j\leq d\}$ to store the variable values before the last sampling of $X_i$.
Hence, for the $i$-th random variable $X_i\in\mathbb{X}$,  we encode the state of its last (re)sampling by the values of $(d_i,(c_{i,1},\dots,c_{i,d}))$.
 
To encode (ii), we introduce $k$ variables $R=\{r_i\bs 1\leq i\leq k \}$ which capture the time  since the last (re)sampling of each random variable $X_i\in\mathbb{X}$ in $r_i$.

Finally, for (iii) we use $\jumpt$ to store the time duration since the last jump. 

Thus our encoding uses $d'=d+k\cdot(d+2)+1$ variables
ordered as $\Var'=\{v_1,\ldots,v_d,d_1,\ldots,d_k,c_{1,1},\ldots,c_{k,d},\allowbreak r_1,\ldots,r_k,\jumpt\}$.
For $\nu\in\R^{d'}$, we use $\nu\proj$ to denote the first $d$ components $(\nu_1,\ldots,\nu_d)$ of $\nu=(\nu_1,\ldots,\nu_{d'})\in\R^{d'}$.
Furthermore, for any $\nu\in\R^{d'}$, $a\in\Var'$ and $b\in\R$, by $\nu[a\mapsto b]$ we denote $\nu$ after changing the entry at the position of variable $a$ (as defined in $\Var'$) to $b$.

 The above encoding is implemented by  extending $\Flow$ to $\Flow'$  in each location $\ell\in\Loc$ with derivative $0$ for each variable in $C \cup D$ and derivative $1$ for each variable in $R\cup\{\jumpt\}$. I.e., for each $\ell\in\Loc$ and $\nu\in\R^{d'}$, $\Flow'(\nu)=(\Flow(\nu\proj),\mathbf{0},\mathbf{1})$ with $\mathbf{0}$ is a sequence of $k(d+1)$ zeros and $\mathbf{1}$ of $k+1$ ones.

Further, $\Edge'{=}\{e'\,|\,e\in\Edge\}$ contains for each $e=(\ell_{j_1},g,r,\ell_{j_2})\in\Edge$ with $\operatorname{a}(e)=i$ the jump $e'=(\ell_{j_1},g',r',\ell_{j_2})$ which extends $e$ to handle the new variables; formally, $g'=\{\nu\in\R^{d'}\,|\,\nu\proj\in g\}$ and for all $\nu\in\R^{d'}$,
$r'(\nu)=\nu[v_1\mapsto r(\nu)[1]]\ldots[v_d\mapsto r(\nu)[d]][d_i\mapsto j_2][c_{i,1}\mapsto r(\nu)[1]]\ldots[c_{i,d}\mapsto r(\nu)[d]][r_i\mapsto 0][\jumpt\mapsto 0]$.

For each $\ell\in\Loc$, $\Init'(\ell)$ consist of all $\nu\in\R^{d'}$ for which $\nu\proj\in\Init(\ell)$, and such that $\nu(d_i)=\ell$, $\nu(c_{i,j})=\nu(v_j)$ and $\nu(r_i)=\nu(\jumpt)=0$ for all $i=1,\ldots,k$ and $j=1,\ldots,d$.

Now we define the kernel $\kernelDelay$. With decomposed scheduling, the duration between two samplings of a random variable (defined by its realisation) might cover consecutive stays in different locations. However, in the composed setting, we are forced to sample a new duration upon entering a new location. 

For each state $\sigma=(\ell,\nu)\in\States_{\mathcal{C}}$ let $\sigma_{X_i}=(\ell_{X_i},\nu_{X_i})\in\States_{\mathcal{D}}$ denote the state in which $X_i$ was (re)sampled the last time, as encoded in the values of the auxiliary variables: $\ell_{X_i}=\ell_{\nu(d_i)}$ and $\nu_{X_i}(v_j)=\nu(c_{i,j})$ for $j=1,\ldots,d$.
For each random variable $X_i\in\mathbb{X}$ with CDF $F_{X_i}^{\sigma_{X_i}}$ and density function $f_{X_i}^{\sigma_{X_i}}$ in state $\sigma_{X_i}$, we first define another random variable $X_i'$ whose probability distribution is first conditioned in that samples are at least as large as the time $\nu(r_i)$ passed since the last (re)sampling of $X_i$, and then shifted by $\nu(r_i)$ to the left:
$F_{X_i'}^{\sigma}(x) = \Pr(X_i\leq \nu(r_i)+x\bs X_i>\nu(r_i))   = \frac{\int_{\nu(r_i)}^{\nu(r_i)+x}f_{X_i}^{\sigma_{X_i}}(t)dt}{1-F_{X_i}^{\sigma_{X_i}}(\nu(r_i))}$.
Let $\mathbb{X}'= \{X_i' \bs X_i \in \mathbb{X} \}$.

For each first state in a (sub)-trace of $\mathcal{D}$ and $\mathcal{C}$, the probability distribution over the delay must be the same in both automata. We let $\kernelDelay$ specify for each state $\sigma\in\States_{\mathcal{C}}$ a probability distribution over  the time delay until the next random variable $X_i\,{\in}\,\mathbb{X}$ expires, i.e.
$\DistDelayKernel=f_M$, where the random variable $M{=}\textit{min}(\mathbb{X}')$ is the minimum over all shifted random variables, as defined in  \cite{arXiv}.

For the kernel $\kernelJump$, we observe that for each random variable $X_i$, in each state $\sigma_\mathcal{D}$ exactly one $X_i$-labeled jump is enabled in $\mathcal{D}$. Our construction of $\Edge'$ maintains this characteristics  in $\mathcal{C}$. To formalize the probability that an enabled jump is taken, we define for each random variable $X_i$ another random variable $X_i''$, $\textbf{X}''=\{X_1'',\ldots,X_k''\}$, which is defined as $X_i'$ but its CDF $F_{X_i''}^{\sigma}(x)$ is shifted with $\nu(r_i)-\nu(\jumpt)$, instead of $\nu(r_i)$, to model the probabilities of samples beyond the time point of the last jump (i.e. in the definition of $F_{X_i'}^{\sigma}(x)$ we replace $\nu(r_i)$ by $\nu(r_i)-\nu(\jumpt)$). We let the discrete kernel define for each state $\sigma_{\mathcal{C}}$ and each edge $e'\in\Edge'$ with $\operatorname{a}(e)=i$ the probability 
\begin{align}
    \kernelJump(\sigma_\mathcal{C})(e')=\begin{cases}
       \Pr^{\sigma_{\mathcal{C}}}_{\mathcal{C}}(X_i''\leq \mathit{min}(\mathbb{X}''\backslash \{X_i''\})) &\text{if $e$ is enabled in $\sigma_\mathcal{C}$, }\\
       0 &\text{otherwise.}
       \end{cases}
\end{align}
Hence, given an arbitrary \HA $\mathcal{D}$ with decomposed scheduling, we can construct a \HA $\mathcal{C}$ with composed scheduling such that $\mathcal{D}$ and $\C$ have the same trace set and the same initial trace set, and such that each common trace  $\tau$ is Pr-equivalent in $\mathcal{D}$  and $\mathcal{C}$.
As furthermore $\kernelDelay$ is specified such that it mimics the distribution over the duration until the next random variable expires, it is assured that $\Pr^\sigma_{\mathcal{D}}(X\leq t)=\Pr^{\sigma}_{\mathcal{C}}(X\leq t)$.

\emph{Statement 2.}
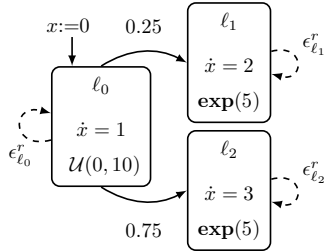
\begin{wrapfigure}[9]{L}{5.1cm}
\centering
\vspace*{-2em}
\resizebox{0.9\linewidth}{!}{%
\begin{tikzpicture}[
	node distance=4.5cm,
	on grid,
	every initial by arrow/.style={-latex},
	thick,
	font=\bfseries\normalsize]

	\node
[
state
] (q0) at (0,0)
{
	$ \ell_0$   \\[10pt]
	$\begin{aligned}
		\dot{x}&=1 \\
	\end{aligned} $ \\[5pt]
	{   $\mathcal{U}(0,10)$}
};

\node[
state
](q1) at (2.2,1.1)
{
	$ \ell_1$   \\[10pt]
	$\dot{x}=2 $  \\[5pt]
	exp$(5)$ 

};
\node[
state
](q2) at (2.2,-1.1)
{
	$ \ell_2$   \\[10pt]
	$\dot{x}=3$ \\[5pt]
	exp$(5)$
};
\path [-latex] 
($(q0.north)+(-0.5,0.5)$) edge [] ($(q0.north)+(-0.5,0)$)
(q0.north) edge [bend left] node [above=0.2, align=center]{$0.25$} ($(q1.west)$)
(q0.south) edge [bend right] node [below=0.25, align=center]{$0.75$} ($(q2.west)$) 
(q0) edge[loop left,dashed,looseness=4] node [below , xshift=0ex,yshift=-5pt,align=center]{$\resamp{\ell_0}$}(q0)
(q1) edge[loop right,dashed,looseness=4] node [above , xshift=2ex,align=center]{$\resamp{\ell_1}$}(q1)
(q2) edge[loop right,dashed,looseness=4] node [above , xshift=2ex,align=center]{$\resamp{\ell_2}$}(q2)
;
\node[draw=none, align=center] at ($(q0.north)+(-0.5,0.7)$) {$ x{:}{=}0 $ };
\end{tikzpicture}
 }   \caption{Counterexample.}
    \label{fig:counterComposed}
\end{wrapfigure} 
To show statement 2 of \Cref{theo:Expressivity} we consider \Cref{fig:counterComposed} as a counterexample. The depicted \HA with composed scheduling is constructed such that in the initial state a delay distributed according to the uniform distribution $\mathcal{U}(0,10)$ is sampled, before a jump to location $\ell_1$ is taken with probability $0.25$ and a jump to location $\ell_2$ with probability $0.75$.

For decomposed scheduling, we  associate each edge with an independent random variable $X_1$ resp. $X_2$, where $\text{min}(X_1,X_2)$ should be uniformly distributed over the intersection of the support of $X_1$ and $X_2$ in location $\ell_0$ to achieve Pr-equivalence. However, the minimum of two continuous random variables with overlapping support can never be uniformly distributed, see \cite{arXiv}.
{\begin{flushright}
\qed
\end{flushright}}

\subsection{Extending (De-)composed Scheduling with Invariants}
\label{subsec:invariants}
If we allow non-trivial invariants, unrealizable time durations might have a positive probability. To manage such cases, we use the concept of \emph{forced jumps} for both, composed and decomposed scheduling. Forced jumps ensure that no time step larger than $\tmax(\sigma)$ is executed.
For both, composed and decomposed scheduling, the \HA is adapted such that each location has a forced jump which is used to  leave a location before its invariant is violated.  
Furthermore, the semantics for composed and decomposed scheduling (\Cref{def:SemanticComposed,def:SemanticDeomposed}) are altered such that for  each state $\sigma$, the time delay is capped by $t'_i=\operatorname{min}(t_i,\tmax(\sigma))$.  

As the probability mass of sampling  a delay larger than $\tmax(\sigma)$ in state $\sigma$ is shifted to the forced jumps, this has to be considered in the probability measures of \Cref{sec:induced} by integrating  over $(\tmax(\sigma),\infty)$ in case of a forced jump.

\section{Classification of Existing Approaches}\label{sec:Relation} 
\FloatBarrier
This paper allows for a broad classification covering multiple formalisms.  Representatives from literature can be found for both variants, as indicated in  \Cref{tab:my_label}. We discuss one representative formalism for composed and one for decomposed scheduling in more depth.
Furthermore, we discuss which formalisms do not fit into  our classification and relate our comparison to the work in \cite{Pola2003StochasticHybridModels} which focuses on Markovian stochastic hybrid models.

\begin{table}[b]
	\caption{Classification of existing formalisms.}
	\label{tab:Classification}
	\centering
	\begin{tabularx}{\textwidth}{|c|*{9}{X|} }
		\cline{2-10}
		\multicolumn{1}{c|}{}                      & \makecell{\cite{Abate2008ProbabilisticSystems}} & \makecell{\cite{Bertrand2014StochasticAutomata}} & \makecell{\cite{Bujorianu2003,Davis1993}} &\makecell{\cite{Delicaris2023Rectangular}}  &{\makecell{\cite{Ghosh1997}}}& \makecell{\cite{Hahn2013ASystems}} & \makecell{\cite{Lygeros2010StochasticApplications}} & \makecell{\cite{Pilch2021OptimizingFlowpipe-Construction,daSilva23}} & \makecell{\cite{Soudjani2013AdaptiveProcesses}} \\ \hline
		\makecell{\bfseries\emph{composed}}   & \makecell{\checkmark}                           &           \makecell{\checkmark}                                           & \makecell{\checkmark}           &&                                            & \makecell{\checkmark}              & \makecell{\checkmark}                               &                                                            & \makecell{\checkmark}                           \\ \hline

		\bfseries \emph{decomposed}           &                                                 &           &                                       & \makecell{\checkmark}                                  & \makecell{\checkmark}                      & \makecell{\checkmark}              &                                                     & \makecell{\checkmark}                                      &                                                 \\ \hline
	\end{tabularx}
	\label{tab:my_label}
\end{table}

\paragraph{Composed Scheduling.}
Lygeros and Prandini \cite{Lygeros2010StochasticApplications} introduce a general class of \emph{continuous-time stochastic hybrid automata} (CTSHA). This approach has been abstracted to discrete-time e.g., in \cite{Abate2008ProbabilisticSystems, Soudjani2013AdaptiveProcesses}.  
CTSHA  implement  composed scheduling, as the stochastic information over the delay is attached to the location and a stochastic kernel chooses the jump-successor state randomly.
 Technically,
 the actual jump times are the stopping time of the inhomogeneous Poisson process with state-dependent rate $\lambda(t) = \lambda(q(t), \textbf{x}(t))$. 
This results in delays which are sampled according to an inhomogeneous exponential distribution which can explicitly be expressed by the stochastic kernel $\kernelDelay$ used for composed scheduling by: $\textit{Pr}\left(X>t \bs \left(\ell(s) ,\mathbf{x}(s)\right )\right)= e^{-\left(\int_{s}^{s+t}\lambda(u)\, du\right)}$.
The integral can be computed, if for each location, the continuous state evolves with a deterministic rate. Thus, $(\ell(s+t'), \textbf{x}(s+t'))$ is well defined for any $t' \leq t \in \Rpz$ and given $(\ell(s), \textbf{x}(s))$.

The inhomogeneous Poisson process used in CTSHA can define a phase-type distribution which can  approximate any continuous probability distribution. Hence, just as for  composed scheduling, the sampled delay  can follow any probability distribution in CTSHA. 
 Moreover, CTSHA directly extend our proposed \HA with  composed scheduling by including stochastic differential equations (SDEs) to describe the  continuous state's evolution. Furthermore, the initial state is sampled according to a probability measure and the kernel $\kernelJump$ is extended to  a continuous stochastic kernel, enabling random resets of the continuous state. %
 
\paragraph{Decomposed Scheduling.}
Pilch et al. \cite{Pilch2021OptimizingFlowpipe-Construction,daSilva23} introduce singular automata with random clocks (SARC) which basically apply decomposed scheduling. They implement this approach by adding \emph{random clocks} which induce random variables characterizing the delay for each  jump. Upon expiration, a new random variable is induced which follows the predefined probability distribution with a constant parameter of the corresponding  random clock. Additionally, SARC allow for transitions which are not associated with the expiration of a random variable which adds non-determinism to the model.
The concept of random clocks can be extended to other sub-classes of hybrid automata, e.g., rectangular automata\cite{Delicaris2023Rectangular} with random clocks, which again implicitly assign random variables to jumps via random clocks and restrict the syntax to ensure a sound probability measure. In contrast, decomposed scheduling allows to directly attach random variables to jumps and the semantics ensures that the resulting probability measure is sound. This simplifies modeling. 
\paragraph{Markovian Stochastic  Hybrid Models.}
 The formalism of \emph{Continuous-Time Mar\-kov Chains} (CTMCs) which are discrete-state models without variables, has been extended in the past to stochastic hybrid models \cite{Davis1993,Bujorianu2003,Ghosh1997,Hu2000TowardsSystems}, which correspond to (different kinds of) Markov  processes.
\emph{Piecewise Deterministic Markov Processes} (PDMPs)  \cite{Davis1993,Bujorianu2003} implement composed scheduling  where the evolution of a continuous state is piece-wise deterministic and can be restricted by invariants. In PDMPs the jump times are ruled by an inhomogenous Poisson process and jumps and their effects are chosen probabilistically by a transition kernel.
This stochastic kernel allows to encode guards implicitly. 

In contrast,  \emph{Switching Diffusion Processes} (SDPs) \cite{Ghosh1997} describe the continuous variables' evolution via stochastic differential equations. They do not allow for invariants or resets of the continuous state and the discrete state evolves according to a \emph{controlled Markov chain}\cite{Kesten1975}, whose transition matrix depends on the continuous state. This allows the user to encode guards into the model. Due to the underlying Markov chain, which can be characterised via a generator matrix describing competing random variables, SDPs can be seen as an approach with decomposed scheduling.  \emph{Stochastic hybrid systems} (SHSs) \cite{Hu2000TowardsSystems} simplify CTSHA as discussed in \cite{Lygeros2010StochasticApplications} by relaxing the inhomogenous Poisson process which determines the random delays. %

The formalisms mentioned above coincide under certain assumptions, as discussed in \cite{Pola2003StochasticHybridModels}. For example, the authors state that the  formalism SDP, which is decomposed in our classification, and the formalism SHS, which we classify as composed, coincide iff invariants and guards of the SHS evaluate to true.

Clearly, restricting to exponentially distributed delays renders the counterexample of  \Cref{theo:Expressivity} invalid, as the minimum of two exponentially distributed random variables is again exponentially distributed. 
However, additional restrictions are necessary to ensure that the probability spaces of (de-)composed scheduling coincide in the presence of guards and invariants. Specifying such restrictions (and proving their correctness) is out of scope for this paper.

 We refer to \cite{Pola2003StochasticHybridModels} for a  more detailed comparison on the expressivity of Markovian stochastic hybrid models.

\paragraph{Other Formalisms.}
Several existing formalisms for stochastic hybrid models do not fit into the proposed classification, as they focus e.g.,  solely on randomly distributed initial states \cite{Shmarov2015ProbReach} or on a non-deterministic choice over discrete probability distributions for choosing a successor state \cite{Sproston2000DecidableAutomata}. 

The formalism presented in \cite{Bertrand2014StochasticAutomata} defines a fully stochastic semantics for timed automata by randomly choosing delays and  jumps. It applies a composed semantics, however without resampling jumps.  Timelock-freedom is ensured by restricting the support of the probability distributions to executable samples.

The formalism presented in \cite{Hahn2013ASystems} proposes networks of stochastic hybrid automata  which fit in both, the  composed as well as the decomposed approach. Such stochastic hybrid automata allow to reset continuous variables to realizations of continuous random variables. Thus, at each jump we can either sample a randomly distributed delay for the location like in composed scheduling, or associate the samples as delays with the jumps as in decomposed scheduling.

\FloatBarrier

\section{Conclusion}
\label{sec:Conclusion}
In this paper we formalized two approaches to extend  hybrid automata with stochasticity.   The first approach  applies \emph{composed scheduling}, where two stochastic kernels are used to sample the  lengths of time steps and the successor states of jumps. In contrast, the second approach yields \emph{decomposed scheduling} via competing random variables. As the realisations of the random variables specify the delay after which the corresponding jump is taken, a race-condition is induced. The minimum realisation of the random variables then fully characterises the next execution step.
We formalized the syntax and semantics for (de-)composed scheduling and  the stochastic processes underlying the different resulting models. We defined \emph{trace probability equivalence} and showed that it is possible to construct for every given \HA with decomposed scheduling, an equivalent \HA with composed scheduling. Via a simple counterexample, we showed that a \HA with composed scheduling exists, for which no equivalent \HA with decomposed scheduling can be constructed. 

To connect the theoretical constructs developed in this paper to existing formalisms, we classified several existing formalisms according to their semantics and pointed to approaches which we cannot capture yet.  To include them in our classification, future work will  consider more expressive systems, e.g, including stochastic resets and  stochastic noise. Furthermore, we plan to investigate the relation of the presented classes to approaches without resampling.

\subsubsection{Acknowledgements.} We thank the ARCH competition 2023 for fruitful discussions on expressing example models from the ARCH report  \cite{ARCH22} in the formalism of Lygeros et al. \cite{Lygeros2010StochasticApplications}.
\newpage
\appendix
\section{Minimum of Two Random Variables}
\label{sec:ProofMinimumUniform}
Let $A$ and $B$ be two independent continuous random variables  with known CDF $F_A(m)$ and $F_B(m)$, as well as known PDF $f_A(m), f_B(m)$.  
We then consider the random variable $M=\operatorname{min}(A,B)$. 
The CDF of the random variable $M$ is given by:
\begin{align*}
        Pr(M \leq m) &= 1-Pr(M > m)\nonumber \\
    &=1-Pr(A>m, B>m) \\
    &= 1-(1-Pr(A\leq m))\cdot(1-Pr(B\leq m))\nonumber\\
    &=1-(1-F_A(m))\cdot(1-F_B(m)) \\
    &= F_A(m)-F_B(m) + F_A(m)F_B(m).
\end{align*}
Hence, the CDF of $M$ is $F_M(m)=F_B(m)-F_A(m) + F_A(m)F_B(m)$. By taking the derivative of  $F_M(m)$ we obtain its PDF $f_M(m):$
\begin{align*}
    f_M(m)&=\frac{\delta F_M(m)}{\delta m}=f_A(m)+f_B(m) - f_A(m)F_B(m) - f_B(m)F_A(m) \\
     &=f_A(m)+f_B(m) - f_A(m)(1-Pr(B\geq m)) - f_B(m)(1-Pr(A\geq m))\\
    &= f_A(m)Pr(B\geq m) + f_B(m)Pr(A\geq m) \\
    &= f_A(m)\int_m^\infty f_B(m) dt + f_B(m)\int_m^\infty f_A(m) dt.
\end{align*}
Hence, the probability density function of $M$ is $f_M(m)=f_A(m)\int_m^\infty f_B(m) dt +f_B(m)\int_m^\infty f_A(m) dt $.

In the following, we require that  $\textit{supp}(f_A))\cap\textit{supp}(f_B) \neq \emptyset$. In the following we show, that $M=\operatorname{min}(A,B)$ cannot be uniformly distributed. 

Rearranging terms we obtain:
\begin{align*}
    f_M(m)&=f_A(m)\int_m^\infty f_B(m) dt + f_B(m)\int_m^\infty f_A(m) dt\\
    &=f_A(m)(1-F_B(m)) + f_B(m)(1-F_A(m)) \\
    &=f_A(m)-f_A(m)F_B(m) + f_B(m)-f_B(m)F_A(m).
\end{align*}

Since $A, B$ are continuous random variables, $f_i(m)\geq 0$, $F_i(m)$  continuous and monotone increasing with $\lim_{m\to\infty}F_i(m) = 1$ and $\lim_{m\to -\infty}F_i(m) = 0$, for $i\in\{A,B\}$. 
Hence, $m_1,m_2 \in \textit{supp}(f_A))\cap\textit{supp}(f_B)$ exist, for which $0\leq F_A(m_1) < F_A(m_2)\leq 1$ and $0 \leq F_B(m_1) < F_B(m_2)\leq 1$ and hence 

\begin{align*}
 F_A(m_1)<F_A(m_2) &\Rightarrow f_B(m_1)F_A(m_1) < f_B(m_2)F_A(m_2)\\
    &\Rightarrow f_B(m_1)-f_B(m_1)F_A(m_1) > f_B(m_2) -f_B(m_2)F_A(m_2), \text{ and }\\
    F_B(m_1)<F_B(m_2) &\Rightarrow f_A(m_1)F_B(m_1) < f_A(m_2)F_B(m_2)\\
    &\Rightarrow f_A(m_1)-f_A(m_1)F_B(m_1) > f_A(m_2) - f_A(m_2)F_B(m_2).
\end{align*}
Thus,  $f_M(m_1)\neq 0  $ and $f_M(m_2)\neq 0$, with  $f_M(m_1) \neq f_M(m_2)$ and therefore $f_M$ cannot be a uniform distribution.
\bibliographystyle{splncs04.bst}
\bibliography{stochasticHybridModels.bib}
\end{document}